\documentclass[useAMS,usenatbib]{mn2e}
\usepackage{amssymb,amsmath} 
\usepackage{graphicx}
\usepackage{epstopdf}
\def \mnras {MNRAS\ }

\title[]{The detection efficiency of on-axis short gamma ray burst optical afterglows triggered by aLIGO/Virgo}

\author[]{D.M. Coward$^{1,2}$\thanks{E-mail:David.Coward@uwa.edu.au}, M. Branchesi$^{3}$, E.J. Howell$^{1}$, P.D. Lasky$^{4}$, M. B\"oer$^{5}$ \\
$^{1}$School of Physics, University of Western Australia, Crawley WA 6009, Australia\\
$^{2}$Australian Research Council Future Fellow\\
$^{3}$DiSBeF - Universit\`a degli Studi di Urbino `Carlo Bo', I-61029 Urbino, Italy\\
$^{4}$School of Physics, University of Melbourne, Parkville, VIC 3010, Australia\\
$^{5}$ Laboratoire ARTEMIS, Observatoire de la C\^{o}te d'Azur, BP 4229, 06304 Nice Cedex 04, France\\
}

\begin{document}
\vspace{-5mm}

\pagerange{\pageref{firstpage}--\pageref{lastpage}} \pubyear{3002}

\maketitle

\label{firstpage}

\begin{abstract}
Assuming neutron star (NS) or neutron star/stellar-mass black hole (BH) mergers as progenitors of the short gamma ray bursts, we derive and demonstrate a simple analysis tool for modelling the efficiency of recovering on-axis optical afterglows triggered by a candidate gravitational wave event detected by the Advanced LIGO and Virgo network.  The coincident detection efficiency has been evaluated for different classes of operating telescopes using observations of gamma ray bursts. We show how the efficiency depends on the luminosity distribution of the optical afterglows, the telescope features, and the sky localisation of gravitational wave triggers. We estimate a plausible optical afterglow and gravitational wave coincidence rate of $1$ yr$^{-1}$ ($0.1$ yr$^{-1}$) for NS-NS (NS-BH), and how this rate is scaled down in detection efficiency by the time it takes to image the gravitational wave sky localization and the limiting magnitude of the telescopes. For NS-NS (NS-BH) we find maximum detection efficiencies of $>80\%$ when the total imaging time is less than $200$ min (80 min) and the limiting magnitude fainter than 20 (21). We show that relatively small telescopes $(m<18)$ can achieve similar detection efficiencies to meter class facilities $(m<20)$ with similar fields of view, only if the less sensitive instruments can respond to the trigger and image the field within 10-15 min. The inclusion of LIGO India into the gravitational wave observatory network will significantly reduce imaging time for telescopes with limiting magnitudes $\sim20$ but with modest fields of view. An optimal coincidence search requires a global network of sensitive and fast response wide field instruments that could effectively image relatively large gravitational-wave sky localisations and produce transient candidates for further photometric and spectroscopic follow-up.
\end{abstract}


\section{Introduction}
Combining electromagnetic (EM) observations with gravitational wave detection is emerging as a significant priority for the upgraded ground-based gravitational wave (GW) detectors \citep{2014ApJS..211....7A,2013arXiv1304.0670L,2012A&A...539A.124L,2012A&A...541A.155A} such as Advanced LIGO
\citep{aligo} and Advanced Virgo (hereafter aLIGO/Virgo) \citep{avirgo}.  It allows GW candidates that are close to the detection threshold to be associated with an optical counterpart that could provide stronger confirmation \citep{cow11}. \citet{koch93,2010ApJ...725..496N} and \cite{schutz} showed that a coincident narrowly beamed gamma ray burst (GRB)-gravitational wave search has the effect of extending the sensitivity horizon distance of GW detectors for a face-on binary neutron star (NS) merger or a NS-stellar black hole (BH) merger, so that the number of potentially detectable GW sources increases by 3.4. The LIGO and Virgo search for GW signals in temporal and spatial coincidence with observed GRBs showed the possibility to study the origin of single events (Abadie et al. 2012a) and improvement in the search sensitivity with respect to the all-sky searches (Abadie et al. 2012b). Multi-messenger observations offer unique opportunities to probe compact objects, such as understanding their birth and evolution and constraining the equation of state of exotic neutron star matter \citep{2013CQGra..30l3001B, 2013arXiv1311.2603B, 2013arXiv1311.1352L}. The comparison between the observed GRB rate and GW rate will provide constraints on the  beaming angle of the GRB progenitor system \citep[e.g.]{2013PhRvL.111r1101C}. 
In the long term, joint observations may also provide a way to determine the luminosity distance--redshift relation: by combining GW derived luminosity distance measurements with EM redshifts, one can potentially constrain key cosmological parameters \citep[e.g.][]{1986Natur.323..310S, 2010CQGra..27u5006S, 2005ApJ...629...15H,2010ApJ...725..496N}.

The most probable EM counterpart of a NS-NS merger is a short gamma ray burst (SGRB), where `short' is defined as $T_{90}<2$s\footnote{The duration in which the cumulative gamma-ray counts increase from 5\% to 95\% above background.}. The favoured model for SGRBs is a compact object merger (NS-NS or NS-BH) triggering an explosion that produces a burst of collimated $\gamma$-rays \citep{elp+89,npp92, lrg05} powered by accretion onto the newly formed compact object. SGRBs are produced by an ultra-relativistic outflow from the central engine with a Lorentz factor of $\Gamma \sim $100--1000.  The outflow is eventually decelerated by interaction with interstellar medium to produce a fading x-ray and optical afterglow (OA).

Merger models predict significant quantities of neutron-rich radioactive species, whose decay should result in a faint transient, known as a `kilonova', \citep{1998ApJ...507L..59L,2005astro.ph.10256K,2010MNRAS.406.2650M} in the days following the burst. Direct evidence for this association has been obtained via deep optical observations of GRB130603B \citep{2013Natur.500..547T, 2013ApJ...774L..23B}, further strengthening the case for the SGRB and neutron star merger model. 


\subsection{GW triggered SGRB afterglow search}\label{rates}
A number of laser interferometric GW detectors reached their
design sensitivities and have been operating as a global array, coordinating
with electromagnetic observations through triggered follow-ups.
These include the LIGO\footnote{http://www.ligo.caltech.edu/}
detectors based at Hanford and Livingston in the USA, the
Virgo\footnote{http://www.virgo.infn.it/} detector in
Italy and the GEO\,600\footnote{www.geo600.uni-hannover.de}
detector in Germany. The LIGO and Virgo detectors are undergoing
a series of upgrades towards Advanced configurations that will produce an order of magnitude improvement in
sensitivity:
aLIGO\footnote{www.ligo.caltech.edu/advLIGO/} and Advanced
Virgo\footnote{www.cascina.virgo.infn.it/advirgo/} are expected to be
operational by 2015.

During the LIGO and Virgo joint science runs in 2009-2010, GW data from three interferometer detectors were analyzed within minutes to select candidate events and infer their apparent sky positions \citep{2012A&A...539A.124L,2012A&A...541A.155A,2014ApJS..211....7A}. The sky positions were sent to several automated telescopes and optical data were obtained for eight such candidates. Although no optical transient counterpart was identified with any of these candidates, and none of the GW triggers showed strong evidence for being astrophysical in nature, the tests proved invaluable in assessing a joint GW and optical search. \cite{2014ApJS..211....7A} evaluated the efficiency for different telescopes and image analysis procedures to detect on-axis optical afterglows. The simulations were performed by injecting over the real images GRB light curves uniformly distributed between the brightest and faintest observed by \cite{kann11}. They showed that short exposures (1 minute) with small aperture telescopes, with observations to depths of less than 18th magnitude, failed to recover SGRB or kilonova light curves at distances comparable to the expected 200 Mpc range of advanced GW detectors to NS-NS mergers. 


There have been numerous studies that attempt to model the EM counterparts of binary compact object mergers and their detection in coincidence with GW searches. Among these, \cite{2012ApJ...746...48M} discuss the joint GW/optical detection for on-axis and off-axis SGRB afterglow models by one-day cadence surveys of wide-field telescopes. They found that the brightest on-axis events (jet energy of $10^{50}$ erg) should be detectable for a few days by surveys such as PTF. \cite{2013ApJ...767..124N} discuss the detectability of isotropic optical counterparts of GW events by telescopes with cameras larger than 1 deg$^2$  and apertures larger than 1 m. They considered a fixed threshold of $\rm M_R < 14 mag$ for at least 2 hr (later relaxed to 11 mag) for NS--NS mergers. The opposite approach (that uses the sky-position and the time of occurrence of SGRBs to perform the GW search) has been recently analyzed by \cite{2013arXiv1308.6081G} for the case of off-axis afterglow detections.


In this paper we focus on face-on compact binary mergers and examine the efficiency of recovering an SGRB OA from a triggered GW search using a ``fast pointing'' observational strategy. We use the SGRBs observations and data (not models) to estimate the merger rate, and the observed optical luminosity function and light curve luminosity dimming of the optical afterglows to provide a realistic estimate for the joint detection efficiency.  We derive a relatively simple formula to estimate the SGRB OA detection efficiency for any telescope, using limiting magnitude, Field of View (FoV), and the telescope observational time to image the whole GW sky localization area.

Specifically, we show how the efficiency for OA detection triggered by a GW event depends on the following factors: 
\begin{itemize}\label{items}
\item Luminosity distribution of SGRB OAs
\item Limiting magnitude (sensitivity) of the imager
\item Exposure time to reach limiting magnitude 
\item Field of view of the CCD
\item Latency to send the GW alert
\item GW sky localisation
\end{itemize}


We excluded from the present analysis off-axis GRBs, for which we do not have observations to constrain luminosity distributions and have to rely only on models. 

The on-axis SGRBs may also be detected in gamma-rays by {\it Swift} with a small localisation uncertainty \citep[see][for a review of {\it Swift} follow-up capabilities]{2012ApJ...759...22K,2012ApJS..203...28E}. However the probability of {\it Swift} detecting an on-axis SGRB is at least a factor 4 smaller than with the Fermi/GBM detector, whose localisation uncertainty is comparable to that of aLIGO/Virgo \citep[see][]{2013ApJ...776L..34S}.

\section{SGRB Optical afterglow selection function}
\subsection{Optical afterglow luminosity function}
One of the problems identified with SGRB OAs triggered by the {\it Swift} satellite is their relative faintness compared to the long GRBs. \cite{kann11} show that SGRB OAs are in general much fainter than long GRBs, up to 5-6 mag dimmer. In addition, a significant fraction do not have any associated OA despite localisation and deep follow-up. This can be partially explained if SGRBs occur in regions where the circum-burst density is relatively low compared to long GRBs, as would be expected if NS mergers are indeed responsible for most SGRBs. There are clearly selection effects that inhibit detection of SGRB OAs, \citep[see][for a study of GRB OA selection effects]{2013MNRAS.432.2141C}. Fortunately, most of the selection effects discussed in \cite{2013MNRAS.432.2141C} occur at high-$z$, so should not bias detection of the more local OAs. Nonetheless, many other non astrophysical biases reduce the efficiency for a SGRB OA detection (see \S \ref{effy}). For example, false candidates are important and indirectly influence the observational strategies (e.g  the choice of multi-color observations, or appropiate resolution), but can be considered more relevant to a post discovery data analysis issue rather than discovery/detection. In this work we ignore these biases and assume that OA detection efficiency depends only on the factors listed in Section \ref{items}.

\begin{figure}
\centering
\includegraphics[scale=0.7]{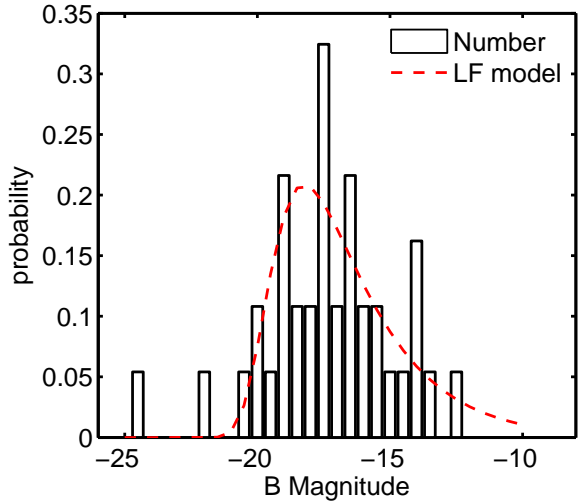}
\caption{The SGRB OA probability distribution function (PDF) using equation \ref{eqn:LF} and GRB OA luminosities from Kann et al. (2011). The optimal model is constrained by the KS test ($P_{\rm{KS}}>0.9$) with $M_0=-19.2$, $\sigma=10$ and $\lambda=0.4$, and normalized over the Magnitude range $[-25,-10]$.  } \label{LF}
\end{figure}

Following \cite{2013MNRAS.432.2141C}, the GRB OA luminosity function (LF) in R-band is approximated by fitting to the compiled SGRB OA luminosities from \cite{kann11}. Luminosities are scaled to 1 day and have been corrected for Galactic extinction. We use the log normal functional form of \cite{Johannesson2007} to approximate the LF: 
\begin{equation}
\label{eqn:LF}
\varphi(L) = C\left(\frac{L}{L_0}\right)^{-\lambda}\exp\left(-\frac{\ln^2(L/L_0)}{2\sigma^2}\right)\exp\left(-\frac{L}{L_0}\right)\;,
\end{equation}
where $C$ is a normalization constant, $L_0$ a characteristic luminosity. We convert $L/L_0$ to absolute magnitudes using $L/L_0=10^{(M_0 - M)/2.5}$. Using the KS test as a constraint ($P_{\rm{KS}}>0.9$), we approximate the LF using $M_0=-19.2$, $\sigma=10$ and $\lambda=0.4$. This LF was converted into a probability distribution function (PDF), plotted in Fig. \ref{LF}, by normalizing over the luminosity range $[-25,-10]$. 

With a reference time $t_\mathrm{c}$ of 1 day and an average OA luminosity decay following $t^{-1}$, the OA limiting luminosity can be calculated at a time $T_\mathrm{i}$ \citep{2013MNRAS.432.2141C} as\footnote{We use the following cosmology in our analysis: $H_0 = 70$ km/s/Mpc, $\Omega_m=0.3$ and $\Omega_{\Lambda}=0.7$.}:

\begin{equation}
\label{eqn:Pogson}
M_\mathrm{L}(z,m_\mathrm{L},T_\mathrm{i}) = m_\mathrm{L} - 5\log_{10}\Big(\frac{d_\mathrm{L}(z)}{10}\Big) - \frac{5}{2} \log_{10}\Big(\frac{T_\mathrm{i}}{t_\mathrm{c}}\Big)\;,
\end{equation}
where $m_\mathrm{L}$ is the limiting magnitude of the telescope.
A SGRB OA dimensionless selection function is obtained by integrating over the OA luminosity distribution:

\begin{equation}
\label{select2}
\psi_{\rm M}(z,m_\mathrm{L},T_\mathrm{i})= \int_{M_\mathrm{L}(z,m_\mathrm{L},T_\mathrm{i})}^{M_\mathrm{Max}}{\varphi(M)} \mathrm{dM}\;.
\end{equation}
where $M_\mathrm{Max}$ is the maximum luminosity (or cut-off) of the luminosity distribution.

\section{SGRB OA externally triggered efficiency function}\label{effy}
We use this model to focus on the detection efficiency of SGRB OAs assuming a telescope is triggered externally via an automated alert from a triple coincidence GW search for NS mergers. 
One significant problem is the size of the localisation uncertainty. This can add significant time to the imaging unless the CCD field of view (FoV) is relatively wide field. Another factor is the  time it takes for a telescope to receive the alert. 

The total imaging time for an externally triggered search is obtained by scaling the time for a single exposure $T_1$ to reach a certain limiting magnitude, by the ratio of the sky area of GW localisation uncertainty $A_{\mathrm{GW}}$ to the telescope (CCD) FoV. This gives a total time, $T_{\mathrm{total}}$ for imaging the entire region:

\begin{equation}
\label{total}
T_{\mathrm{total}}= { T_{1}} \times\frac{A_{\mathrm{GW}}}{FoV} + T_{\mathrm{GW}}\;,
\end{equation}
\noindent where $T_{\mathrm{GW}}$ is the time it takes to send a GW alert from the trigger time.

Defining $R$ as the coincident rate of GW/SGRB events and $\rho$, the rate density of NS-NS mergers inside the GW detection volume (see below), then $R=(\rho/B)V$, where $B$ is the beaming factor. Hence the dimensionless optical and GW coincidence detection efficiency as a function of imaging time and telescope limiting mag is defined as:


\begin{equation}
\label{efficiency}
\mathrm {Eff}(m_\mathrm{L},T_{\mathrm{total}}) =\frac{\int_{0}^{0.07} dR/dz \;\psi_{\rm M}(z,m_\mathrm{L},T_{\mathrm{total}}) \; \mathrm{dz}}{\int_{0}^{0.07} dR/dz \; \mathrm{dz}},
\end{equation}
where $dR/dz$ is the differential rate. For the total rate we use $z=0.07$ (300 Mpc), corresponding to the aLIGO/Virgo NS merger detection range (location and orientation average distance, 197 Mpc), multiplied by a factor 1.5 to take into account the stronger gravitational wave emission for face on mergers \citep{schutz}.

\begin{figure*}
\includegraphics[scale=0.78]{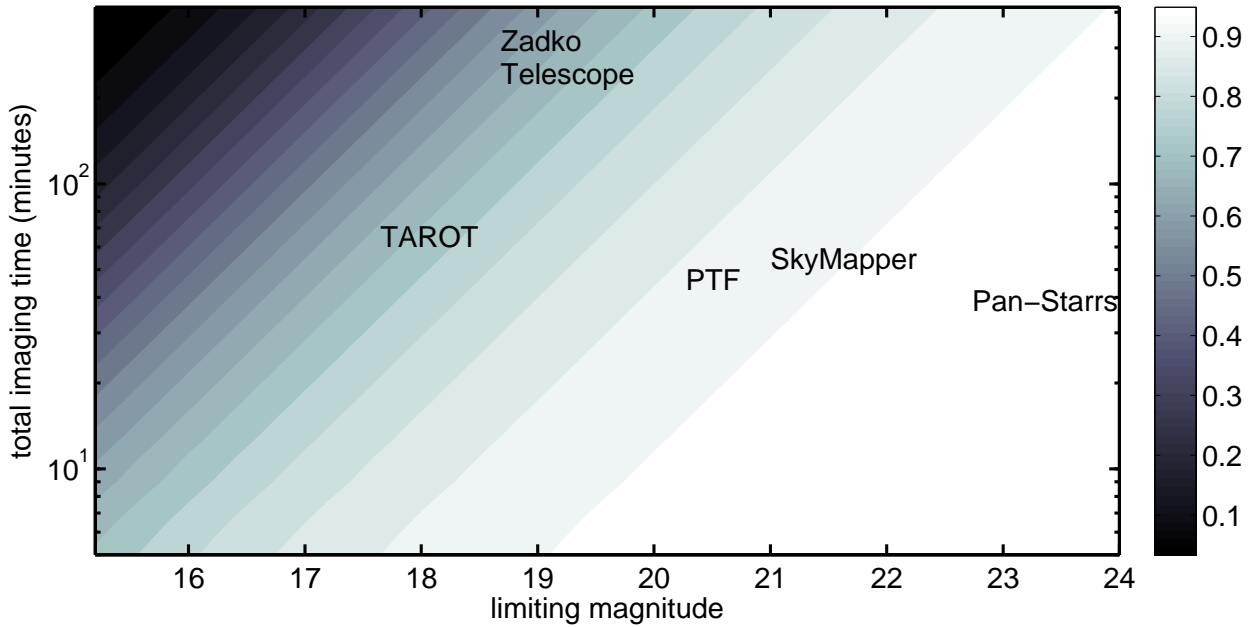}
\caption{Density plot showing the detection rate (number per year) of SGRB OAs (gray scale) in the imaging time versus telescope limiting magnitude plane for NS-NS mergers. Using Table 1, we have inserted several telescope specifications encompassing a range of FoVs and limiting magnitudes. The total imaging time for each instrument depends strongly on the CCD FoV. For total imaging times less than 200 min, we find that the sensitive wide FoV facilities, such as Pan-STARRS, PTF and SkyMapper, can detect a SGRB OA at a rate of 0.8 yr$^{-1}$. With the inclusion of LIGO India into the network, there is a localisation uncertainty reduction by a factor of 3.7. The most significant detection rate improvement (18\%) occurs for the Zadko Telescope, but the improvement for the wide FoV facilities is less than $(1-3)$\%. See also Table 2. for telescope specific detection rates for NS-NS and NS-BH mergers.} 
\label{fig_eff} 
\end{figure*}

 Assuming a significant fraction of binary neutron star mergers produces SGRBs, \cite{2012MNRAS.425.2668C} calculate a 
detection rate of (0.2-40) yr$^{-1}$ for the aLIGO and Virgo binary NS range \citep[see][for detection ranges]{2013arXiv1304.0670L,abadie2010}. For a joint SGRB OA and GW search using 
aLIGO and Virgo, the above rates are scaled down to account for the beaming factor of SGRB OAs. Assuming a beaming 
half angle of 8 degree \citep{2013arXiv1309.7479F}, or a beaming factor of about 100, the coincident optical and aLIGO/Virgo detection rate is about 1 yr$^{-1}$ assuming a 
100\% efficiency for a 300 Mpc radius volume. The efficiency function can be used to scale this rate using a total imaging time and limiting magnitude that corresponds to any given instrument. 
Figure \ref{fig_eff} shows a density plot of equation (\ref{efficiency}), as a function of total imaging time and telescope limiting magnitude. 

Following the suggestion of \cite{2008MNRAS.385L..10T}, which identify NS-BH mergers as possible progenitors of the extended emission SGRB, \cite{2012MNRAS.425.2668C} estimated a merger rate of 0.16--7.1 Gpc$^{-3}$ yr$^{-1}$ (consistent with NS-BH merger rate of \cite{abadie2010} and \cite{2013ApJ...779...72D}). The corresponding GW detection rate is 0.05-- 2.0 yr$^{-1}$ for the aLIGO and Virgo NS-BH (415 Mpc) range. Following the same assumptions used for NS-NS and considering a search volume of 615 Mpc radius, we obtained the detection efficiencies for a SGRB OA and a coincident optical and aLIGO/Virgo rate of 0.07 yr$^{-1}$. Even if the NS-BH detections are expected to be less likely, their observations would shed light on the dynamics of the binary system, on its parameters, and on the internal structure of its components as shown by \cite{2014PhRvD..89f4056M}. 

The efficiency function (above) is easily converted to a telescope specific detection rate, $R_T$, by the product $R_T=\mathrm {Eff}(m_\mathrm{L},T_{\mathrm{total}}) R$, where $R$ is the OA and GW coincidence rate of $1$ yr$^{-1}$ ($0.1$ yr$^{-1}$) for NS-NS (NS-BH) respectively.
   
\begin{table}
 \begin{tabular}{@{}lccccc}
\hline
\hline
Telescope & limiting & Exposure & FoV & Total imaging$^\dagger$ \\
& mag & (s) & deg$^2$ & time (min) \\
\hline
Zadko  & 21 & 180 & 0.15 & 370$^\ddagger$ \\
TAROT & R $\approx 18$ & 60 & 3.5 & 44\\
SkyMapper & g $\approx  21.9$ & 110 & 5.7 & 62\\
PTF 1.2m & R $\approx 20.6$ & 60 & 8.1 & 42 \\
Pan-STARRS & R $\approx 24$ & 30 & 7.0 & 37 \\
\hline
\hline

\end{tabular}
\caption[]{We employ the following telescopes, Zadko Telescope \citep{cow10}, TAROT \citep{2013Msngr.151....6K}, SkyMapper \cite{2007PASA...24....1K}, PTF \citep{2009AAS...21346901L} and Pan-STARRS \citep{2009AAS...21330107C}, with different FoV and limiting sensitivities used for calculating coincident detection efficiencies. Total imaging time for each telescope employs equation \ref{total} and for definiteness we use a 100 square degree localisation uncertainty, equal probability of detection across the entire region, and a time to receive the GW alert, $T_{\mathrm{GW}}$, of 30 min.\newline
 $^\dagger$ We do not include other factors that also contribute to the total imaging time, such as telescope slewing time and CCD readout time.\newline
  $^\ddagger$ For Zadko the total imaging time is based on $m=19$ for a $0.5$ minute exposure. }
\end{table}\label{table1}

For definiteness, we have included the expected efficiencies for several operating telescopes that could be used in a optical and GW coincidence search (see Table 1 for a list of telescope characteristics).
For the telescopes listed in Table 1, we assume that $A_{\mathrm{GW}}$ is geometrically simple, i.e. a square or rectangle and there is an equal probability of the optical source being located across the entire region of $A_{\mathrm{GW}}$, so that $T_{\mathrm{total}}$ defines the maximum time for source identification. In reality, the GW localisation uncertainty is non-uniform, and the sky localisation regions are often geometrically complex. We do not consider this affect on imaging time here, but focus on how limiting magnitude, imaging time, and localisation uncertainty regardless of geometry, are related and how they define a flux limited detection efficiency.

In this study we use a conservative 100 square degree localisation uncertainty for the aLIGO/Virgo network, and following \cite{ligop}, a localisation improvement of a factor of 3.7 by including LIGO India in the network. The GW alert time is fixed at 30 min on the basis of the 2009-2010 optical follow-up campaign \citep{2014ApJS..211....7A}.

\section{Results and discussion}
We derive and demonstrate a simple analysis tool to constrain the efficiency of detecting a SGRB OA associated with a GW candidate alert. The sensitivity, FoV and fast reaction of the telescopes are important factors for the recovery of a coincident SGRB OA and a NS merger candidate triggered by aLIGO/Virgo. Our main results are shown shown in figure 2, which plots the coincident detection efficiency  for a NS-NS merger as contours in the total imaging time versus limiting magnitude plane. Maximum efficiencies are obtained from a combination of wide FoV and high sensitivities, where the total imaging time is a minimum. Fixing the GW alert latency to 30 minutes, we find detection efficiencies $>80\%$ when the total telescope imaging time is in the range $(35-200)$ min at a limiting magnitude range of $(20-24)$ respectively. Reducing the GW alert latency to 5 minutes, we found that a small telescope with a limiting magnitude of 18 and a FoV of 10 sq. degree and fast pointing strategy (15 min total imaging time), has equivalent efficiency to a meter class telescope with limiting magnitude 21 with a 240 min total imaging time.

Table 2 shows the OA and GW coincidence rate for NS-NS and NS-BH. It also shows that the inclusion of LIGO India into the GW observatory network will significantly increase the detection efficiency for telescopes with limiting magnitudes $\sim20$ but with modest FoVs.

\begin{table}
 \begin{tabular}{@{}lcccc}
\hline
\hline
Telescope & $R_T$ Yr$^{-1}$ & $R_T$ Yr$^{-1}$ \\
& NS-NS (NS-BH) & NS-NS (NS-BH) \\
& aLIGO/Virgo & aLIGO/Virgo/LIGOI\\ 
\hline
Zadko  & 0.62 (0.037) & 0.76 (0.058)  \\
TAROT & 0.77 (0.06) & 0.8 (0.064) \\
SkyMapper & 0.95 (0.09) & 0.96 (0.092) \\
PTF 1.2m & 0.92 (0.087) & 0.93 (0.086)  \\
Pan-STARRS & 0.98 (0.097) & 0.98 (0.097)  \\
\hline
\hline

\end{tabular}
\caption[]{Telescope specific detection rates using plausible SGRB OA and GW coincidence rate of $1$ yr$^{-1}$ ($0.1$ yr$^{-1}$) for NS-NS (NS-BH) triggered by a triple coincidence aLIGO/Virgo, and (second column), a networked GW search with inclusion of LIGO India. For the latter network, the sky localisation is improved by a factor 3.7 \citep{ligop}, and the total imaging time is reduced accordingly. The highest detection rate improvement occurs for the Zadko Telescope, while the wide field facilities are more limited by the GW alert time (fixed at 30 min). }
\end{table}\label{table2}

As shown by \cite{2014ApJS..211....7A}, there are other factors that can significantly reduce the efficiency for an optical counterpart to be identified with a GW trigger. These factors include non-optimal observing conditions, image quality, and sky localisation in crowded star fields or bright galaxies.
Furthermore,  the detection of a unique optical counterpart over hundreds of square degrees sky area deals with many thousands of contaminant transients (cosmic rays, asteroids, variable stars, M-dwarf flares, Active Galactic Nuclei variability, supernovae, etc.)
The removal of contaminants requires the use of multi-color observations of an appropriate cadence, multi-epoch observations, and the development of data analysis procedures able to quickly classify the transients \citep[see e.g.][]{2014ApJS..211....7A,2013ApJ...779...18B,2013ApJ...767..124N}. The discovery of a GRB afterglow in 71 square degrees by \cite{2013ApJ...776L..34S} proved the practicality to identify faint transients searching over a wide sky area.
 
This paper takes into account the small rate of on-axis SGRB, their observed luminosity function, and their fast luminosity decay, and it highlights and quantifies the importance of a fast reaction of optical telescopes to obtain at least a first observation of the entire GW error box, which is necessary to identify the optical counterpart of the GW source.
Our choice of telescopes for the EM follow-up does not cover all the facilities for the aLIGO/Virgo era, but is representative of different classes of instruments that are currently operating with a fast pointing capability. There are projects for larger FoV facilities, such as BlackGEM, GOTO, iPTF and the Large Synoptic Survey Telescope. The latter two will operate in survey mode. 
Another potentially optimal telescope, if it will operate in a rapid response mode, is the 2.6m VLT survey telescope: its sensitivity, large FoV (1 sq. degree), and UV- to I- observation bands will provide deep multicolor observations of candidate mergers. 

Even though there is a clear detection efficiency bias towards sensitive wide-field instruments, there also exists a significant geographical bias in the distribution of automated follow-up telescopes. There are more automated telescopes in the Northern Hemisphere, compared to the Southern Hemisphere. This bias is evidenced by the sky distribution of {\it Swift} triggered GRB OAs, where there is a bias against detecting OAs at very negative declinations. A telescope network with even longitude distribution will be important to sample the light curve over hour long time-scales, so that the source can be unambiguously identified as the GW optical counterpart and not a background/forground transient. Hence, a reliance on a relatively small number of wide FoV telescopes coupled with an expected small OA-GW coincident rate of $1$ yr$^{-1}$, implies that global sky coverage becomes critical. A distribution of fully robotic telescopes, such as the TAROT network and Zadko Telescope \citep{cow10}, will help to fill the observational gaps in latitude/longitude for optimal sky coverage. 

Our work shows that an optimal observation strategy implies a low latency GW alert (a few minutes), a fast telescope reaction, automated image acquisition and a global telescope network.
For the present network, our analysis support the case for a hierarchical type search, as discussed in \cite{2014ApJS..211....7A}, where the sensitive wide FoV instruments can effectively image the largest fields and produce transient candidates for further follow-up. 


\section*{Acknowledgments}
D.M. Coward is supported by an Australian Research Council Future Fellowship (FT100100345). E.J. Howell acknowledges support from a University of Western Australia Fellowship. M Branchesi is supported by the Italian Ministry of Education, University and Research via grant FIRB 2012-RBFR12PM1F. P.D. Lasky is supported by the Australian Research Council Discovery Project (DP110103347).

\label{lastpage}

\begin{thebibliography}{99}

\bibitem[Aasi et al.(2014)]{2014ApJS..211....7A} Aasi, J., Abadie, J., Abbott, B.~P., et al.\ 2014, ApJS, 211, 7 

\bibitem[\protect\citeauthoryear{Aasi et 
al.}{2013b}]{2013arXiv1304.0670L} Aasi J., et al., 
2013b, (arXiv:1304.0670)

\bibitem[\protect\citeauthoryear{Abadie et al.}{2010}]{abadie2010}Abadie J. et al. 2010 Class. Quant. Grav., 27, 173001

\bibitem[\protect\citeauthoryear{Abadie et al.}{2012a}]{2012ApJ...755...2A} 
Abadie J., et al., 2012a, ApJ, 755, 2

\bibitem[\protect\citeauthoryear{Abadie et al.}{2012b}]{2012ApJ...760...12A} 
Abadie J., et al., 2012b, ApJ, 760, 12

\bibitem[\protect\citeauthoryear{Abadie et al.}{2012c}]{2012A&A...539A.124L} Abadie J. et al., 2012c, A\&A, 539, A124

\bibitem[\protect\citeauthoryear{Abadie et 
al.}{2012d}]{2012A&A...541A.155A}Abadie J., et al., 2012d, A\&A, 541, A155


%
%

\bibitem[\protect\citeauthoryear{Acernese et al.}{2009}]{avirgo} Acernese F. et al., 2009, Advanced Virgo baseline design, \textit{Virgo Internal Note} VIR-0027A-09


\bibitem[\protect\citeauthoryear{Bartos, Brady, 
\& M{\'a}rka}{2013}]{2013CQGra..30l3001B} Bartos I., Brady P., M{\'a}rka S., 2013, Class. Quant. Grav., 30, 123001

\bibitem[\protect\citeauthoryear{Berger}{2013}]{2013arXiv1311.2603B} Berger 
E., 2013, arXiv, arXiv:1311.2603

\bibitem[Berger et al.(2013)]{2013ApJ...779...18B} Berger E., Leibler, C.~N., Chornock, R., et al.\ 2013, ApJ, 779, 18 

\bibitem[\protect\citeauthoryear{Berger, Fong, 
\& Chornock}{2013}]{2013ApJ...774L..23B} Berger E., Fong W., Chornock R., 2013, ApJ, 774, L23

\bibitem[\protect\citeauthoryear{Chambers}{2009}]{2009AAS...21330107C} 
Chambers K.~C., 2009, AAS, 41, \#301.07

\bibitem[\protect\citeauthoryear{Chen 
\& Holz}{2013}]{2013PhRvL.111r1101C} Chen H.-Y., Holz D.~E., 2013, PRL, 111, 181101

\bibitem[\protect\citeauthoryear{Coward et al.}{2010}]{cow10}Coward D. M., et al. 2010, Publications of the Astronomical Society of Australia, 27, 331 


\bibitem[Coward et al.(2011)]{cow11}Coward D. M. et al., 2011, MNRAS 415, L26

\bibitem[\protect\citeauthoryear{Coward et al.}{2012}]{2012MNRAS.425.2668C} 
Coward D.~M., et al., 2012, MNRAS, 425, 2668

\bibitem[\protect\citeauthoryear{Coward et al.}{2013}]{2013MNRAS.432.2141C} 
Coward D.~M., Howell E.~J., Branchesi M., Stratta G., Guetta D., Gendre B., 
Macpherson D., 2013, MNRAS, 432, 2141

\bibitem[\protect\citeauthoryear{Dalal et al.}{2006}]{2006PhRvD..74f3006D} 
Dalal N., Holz D.~E., Hughes S.~A., Jain B., 2006, PhRvD, 74, 063006

\bibitem[\protect\citeauthoryear{Dominik et 
al.}{2013}]{2013ApJ...779...72D} Dominik M., Belczynski K., Fryer C., Holz 
D.~E., Berti E., Bulik T., Mandel I., O'Shaughnessy R., 2013, ApJ, 779, 72

\bibitem[Eichler et al.(1989)]{elp+89} Eichler D. et al.\ 1989, Nature, 340, 126

\bibitem[\protect\citeauthoryear{Evans et al.}{2012}]{2012ApJS..203...28E} 
Evans P.~A., et al., 2012, ApJS, 203, 28

\bibitem[\protect\citeauthoryear{Fong et al.}{2013}]{2013arXiv1309.7479F} 
Fong W.-f., et al., 2013, arXiv:1309.7479

\bibitem[\protect\citeauthoryear{Ghosh 
\& Bose}{2013}]{2013arXiv1308.6081G} Ghosh S., Bose S., 2013, arXiv, arXiv:1308.6081 

\bibitem[\protect\citeauthoryear{Harry et al.}{2010}]{aligo} Harry G.~M. et al., 2010, Class. Quant. Grav., 27, 084006

\bibitem[\protect\citeauthoryear{Holz 
\& Hughes}{2005}]{2005ApJ...629...15H} Holz D.~E., Hughes S.~A., 2005, ApJ, 629, 15

\bibitem[J\'{o}hannesson, Bj\"{o}rnsson \& Gudmundsson(2007)]{Johannesson2007}J\'{o}hannesson G., Bj\"{o}rnsson G., Gudmundsson D.H., 2007, A\&A, 472, L29

\bibitem[\protect\citeauthoryear{Kann et al.}{2011}]{kann11}
Kann D.~A., et al. 2011, ApJ, 734, 96

\bibitem[\protect\citeauthoryear{Kanner et al.}{2012}]{2012ApJ...759...22K} 
Kanner J., Camp J., Racusin J., Gehrels N., White D., 2012, ApJ, 759, 22


\bibitem[\protect\citeauthoryear{Keller et al.}{2007}]{2007PASA...24....1K} 
Keller S.~C., et al., 2007, Publications of the Astronomical Society of Australia, 24, 1


\bibitem[\protect\citeauthoryear{Klotz et al.}{2013}]{2013Msngr.151....6K} 
Klotz A., Boer M., Atteia J.-L., Gendre B., Le Borgne J.-F., Frappa E., 
Vachier F., Berthier J., 2013, Msngr, 151, 6

\bibitem[Kochanek \& Piran (1993)]{koch93} Kochanek C. S., Piran, T. 1993, ApJ, 417, L17



\bibitem[\protect\citeauthoryear{Kulkarni}{2005}]{2005astro.ph.10256K} 
Kulkarni S.~R., 2005, astro, arXiv:astro-ph/0510256

\bibitem[\protect\citeauthoryear{Lasky et al.}{2013}]{2013arXiv1311.1352L} 
Lasky P.~D., Haskell B., Ravi V., Howell E.~J., Coward D.~M., 2013, (arXiv:1311.1352)


\bibitem[\protect\citeauthoryear{Law et al.}{2009}]{2009AAS...21346901L} 
Law N.~M., Kulkarni S., Ofek E., Quimby R., Kasliwal M., Palomar Transient 
Factory Collaboration, 2009, AAS, 41, \#469.01

\bibitem[Lee et al.(2005)]{lrg05} Lee W.~H., Ramirez-Ruiz, E., Granot, J.\ 2005, ApJL, 630, L165

\bibitem[\protect\citeauthoryear{Li 
\& Paczy{\'n}ski}{1998}]{1998ApJ...507L..59L} Li L.-X., Paczy{\'n}ski B., 1998, ApJ, 507, L59

\bibitem[\protect\citeauthoryear{Maselli 
\& Ferrari}{2014}]{2014PhRvD..89f4056M} Maselli A., Ferrari V., 2014, PhRvD, 89, 064056

\bibitem[\protect\citeauthoryear{Metzger et 
al.}{2010}]{2010MNRAS.406.2650M} Metzger B.~D., et al., 2010, MNRAS, 406, 
2650


\bibitem[\protect\citeauthoryear{Metzger 
\& Berger}{2012}]{2012ApJ...746...48M} Metzger B.~D., Berger E., 2012, ApJ, 746, 48

\bibitem[Narayan et al.(1992)]{npp92} Narayan R., Paczynski B., Piran, T.\ 1992, ApJL, 395, L83

\bibitem[\protect\citeauthoryear{Nissanke et 
al.}{2010}]{2010ApJ...725..496N} Nissanke S., Holz D.~E., Hughes S.~A., 
Dalal N., Sievers J.~L., 2010, ApJ, 725, 496

\bibitem[\protect\citeauthoryear{Nissanke, Kasliwal, 
\& Georgieva}{2013}]{2013ApJ...767..124N} Nissanke S., Kasliwal M., Georgieva A., 2013, ApJ, 767, 124 

\bibitem[\protect\citeauthoryear{Nissanke et 
al.}{2013}]{2013arXiv1307.2638N} Nissanke S., Holz D.~E., Dalal N., Hughes 
S.~A., Sievers J.~L., Hirata C.~M., 2013, (arXiv:1307.2638)

\bibitem[\protect\citeauthoryear{Sathyaprakash, Schutz, 
\& Van Den Broeck}{2010}]{2010CQGra..27u5006S} Sathyaprakash B.~S., Schutz B.~F., Van Den Broeck C., 2010, Class. Quant. Grav., 27, 215006

\bibitem[\protect\citeauthoryear{Sathyaprakash et al.}{2013}]{ligop}Sathyaprakash B. et al., 2012, LIGO public document, T1200219-v1 

\bibitem[\protect\citeauthoryear{Schutz}{1986}]{1986Natur.323..310S} Schutz 
B.~F., 1986, Nature, 323, 310

\bibitem[Schutz(2011)]{schutz}Schutz B., 2011, Class. Quant. Grav., 28,125023

\bibitem[\protect\citeauthoryear{Singer et al.}{2013}]{2013ApJ...776L..34S} 
Singer L.~P., et al., 2013, ApJ, 776, L34

\bibitem[\protect\citeauthoryear{Tanvir et al.}{2013}]{2013Natur.500..547T} 
Tanvir N.~R., Levan A.~J., Fruchter A.~S., Hjorth J., Hounsell R.~A., 
Wiersema K., Tunnicliffe R.~L., 2013, Nature, 500, 547

\bibitem[Troja et al.(2008)]{2008MNRAS.385L..10T} Troja E., King, A.~R., 
O'Brien, P.~T., Lyons, N., \& Cusumano, G.\ 2008, \mnras, 385, L10 


\end{thebibliography}
\end{document}